\documentclass[11pt,a4paper]{article}

\usepackage{jheppub}						
\usepackage[T1]{fontenc}					
\usepackage{mathtools,physics} 				
\usepackage{graphicx}						
\usepackage{subcaption} 			
\usepackage{cleveref}						
\usepackage{orcidlink}						

\crefname{figure}{figure}{figures}

\preprint{
	\begin{minipage}{5cm}
		\small
		\flushright
		KYUSHU-HET-319\\
		KANAZAWA-25-01
	\end{minipage}}

\title{Pseudo-Nambu-Goldstone-boson Dark Matter from Three Complex Scalars}

\author[a]{Riasat Sheikh\,\orcidlink{0009-0007-1207-1358},}
\author[b,c]{Takashi Toma\,\orcidlink{0000-0001-5828-0090},}
\author[a,d]{and Koji Tsumura\,\orcidlink{0000-0003-3765-2750}}

\affiliation[a]{Department of Physics, Kyushu University,\\ 744 Motooka, Nishi-ku, Fukuoka, 819-0395, Japan}
\affiliation[b]{Institute of Liberal Arts and Science, Kanazawa University,\\ Kanazawa 920-1192, Japan}
\affiliation[c]{Institute for Theoretical Physics, Kanazawa University,\\ Kanazawa 920-1192, Japan}
\affiliation[d]{Kavli IPMU (WPI), UTIAS, University of Tokyo,\\ Kashiwa, 277-8584, Japan}

\emailAdd{riasat.sheikh@phys.kyushu-u.ac.jp}
\emailAdd{toma@staff.kanazawa-u.ac.jp}
\emailAdd{tsumura.koji@phys.kyushu-u.ac.jp}

\date{\today}

\abstract{This study explores a dark matter model in which a pseudo-Nambu-Goldstone boson arises as a viable dark matter candidate from the spontaneous and soft breaking of global $\mathrm{U}(1)$ symmetries and stabilized by a residual $\mathbb{Z}_3$ discrete symmetry. The model introduces three complex scalar fields, singlets under the Standard Model gauge group, and charged under a dark $\mathrm{U}(1)_V$ gauge symmetry together with a permutative exchange symmetry among three scalars. These features naturally suppress the dark matter--nucleon scattering cross section by its Nambu-Goldstone boson nature. In addition to conventional annihilation channels, the $\mathbb{Z}_3$ structure allows semi-annihilation processes, playing a crucial role in setting the relic abundance.
	We analyze theoretical and experimental constraints, including relic abundance, Higgs invisible decays, and perturbative unitarity, and
	evaluate the elastic scattering cross section for boosted dark matter.
}


\begin{document}
\maketitle
\flushbottom

\section{Introduction}
\label{sec:intro}

The existence of dark matter (DM) is one of the most compelling pieces of evidence for physics beyond the Standard Model (SM). The gravitational effects of DM are well-established, as evidenced by the rotation curves of galaxies and the large-scale structure of the universe. However, its particle nature remains elusive, with no direct detection of DM particles to date \cite{PandaX-4T:2021bab,LZ:2022lsv,XENON:2023cxc}. Among several viable candidates, the weakly interacting massive particle (WIMP) scenario has long served as a benchmark due to its natural consistency with thermal freeze-out production. However, null results from direct detection experiments have imposed increasingly stringent constraints on WIMP models, motivating the search for alternative mechanisms that naturally evade such limits.

A compelling class of candidates arises from pseudo-Nambu-Goldstone boson (pNGB), which emerge from the spontaneous and soft breaking of symmetries \cite{Gross:2017dan}. Models for the pNGB DM are particularly attractive as their derivative-dominated interactions (in a non-linear representation of scalars) suppress scattering amplitudes at low momentum transfer, thereby remaining consistent with the latest bounds from direct detection experiments \cite{PandaX-4T:2021bab,LZ:2022lsv,XENON:2023cxc}. At the same time, these models retain sufficient annihilation cross section of the DM into SM particles  i.e., $\expval{\sigma v_\text{rel}} \simeq 10^{-26}\,\text{cm}^3 \text{s}^{-1}$ to account for the observed relic density $\Omega_{\text{DM}} h^2 = 0.12 \pm 0.001$ \cite{Planck:2018vyg}.

The original pNGB model proposed in~\cite{Gross:2017dan} successfully addressed the direct detection problem but suffered from the domain wall (DW) issue (See also~\cite{Karamitros:2019ewv}). In the subsequent extended models~\cite{Abe:2020iph,Okada:2020zxo,Abe:2021byq,Okada:2021qmi,Abe:2022mlc,Liu:2022evb,Otsuka:2022zdy,Abe:2024vxz,Abe:2024lzj} (See also ~\cite{Abe:2021nih,Abe:2021vat,Cai:2021evx,Cho:2023hek,Maji:2023fba}), the DW problem was avoided by embedding the softly broken symmetry into a gauge symmetry while preserving the key features of the pNGB framework.

In this work, we propose a new pNGB DM model based on three SM-singlet complex scalar fields charged under three dark $\mathrm{U}(1)_V$ gauge symmetry. The spontaneous breaking of these symmetries, along with a softly broken $\mathrm{U}(1)_A$ global symmetry, leads to the emergence of pNGBs. A residual $\mathbb{Z}_3$ symmetry stabilizes the DM candidate and enables semi-annihilation processes. These processes are forbidden in $\mathbb{Z}_2$-based models \cite{Abe:2024vxz} and give rise to a new contribution to set the dark matter relic abundance. In particular, they can produce highly energetic, or boosted dark matter (BDM) particles that may be detectable at neutrino and direct detection experiments. One of the motivations for this model is to provide a concrete theoretical framework for realizing the semi-annihilation-driven pNGB BDM scenario investigated in \cite{Aoki:2023tlb, Toma:2021vlw, Miyagi:2022gvy, BetancourtKamenetskaia:2025noa}, and to explore whether such processes can leave observable imprints in current or future detectors. However, as shown later in this work, the resulting BDM flux is not sufficient for detection in the present setup.

This paper is organized as follows. In \cref{sec:model}, we introduce the model and its Lagrangian, including the scalar potential and gauge kinetic mixing.
We also discuss the mass spectrum and the parameters of the model.
In \cref{sec:constraints}, we analyze the constraints on the model parameters, including perturbative unitarity, Higgs invisible decay.
In \cref{sec:DM}, we study annihilation channels, as well as the semi-annihilation processes that can lead to BDM particles. We also explore the allowed parameter space consistent with thermal relic abundance of DM.
Finally, we conclude in \cref{sec:conclusion}.

\section{The Model}
\label{sec:model}

\subsection{Defining the Lagrangian
}

We introduce three complex scalars $S_1,S_2$ and $S_3$ which are SM singlets and transform  under a gauged $\mathrm{U}(1)_V$ symmetry, as

\begin{equation}
	S_1 \to e^{i \theta_V(x)} S_1, \qquad
	S_2 \to e^{i \theta_V(x)} S_2, \qquad
	S_3 \to e^{i \theta_V(x)} S_3,
\end{equation}
where $\theta_V^{}(x)$ is the real-valued spacetime dependent gauge parameter. We also impose a softly-broken global $\mathrm{U}(1)_A$ symmetries, under which the fields transform as
\begin{equation}
	S_1 \to e^{i \theta_A^1} S_1, \qquad
	S_2 \to e^{i \theta_A^2} S_2, \qquad
	S_3 \to e^{i \theta_A^3} S_3,
\end{equation}
where $\theta_A^a\,(a=1,2,3)$ are the relative rotation phases. In addition, we introduce an exact discrete $S(3)$ permutative exchange symmetry among the complex scalars:
\begin{equation}
	S_1 \leftrightarrow S_2 \qquad S_2 \leftrightarrow S_3 \qquad S_3 \leftrightarrow S_1.
\end{equation}

The Lagrangian of our model is given as
\begin{align}\label{eq7}
	\mathcal{L} & = \mathcal{L}_{\text{SM} \, \text{(w/o Higgs potential)}} + \abs{D_{\mu} S_1}^2 + \abs{D_{\mu} S_2}^2 + \abs{D_{\mu} S_3}^2
	\nonumber                                                                                                                                 \\
	            & \qquad - \frac{1}{4} V^{\mu \nu} V_{\mu \nu} - \frac{ \sin \epsilon}{2} V^{\mu \nu} Y_{\mu \nu}
	- \mathcal{V}(S_1, S_2, S_3, \Phi),
\end{align}
where $\epsilon$ is the kinetic mixing angle, and the scalar potential in \cref{eq7} of our model is given by
\begin{align}\label{eq8}
	\mathcal{V}(S_1, S_2, S_3, \Phi)
	 & = \mu_S^2 \qty\Big(\abs{S_1}^2 + \abs{S_2}^2 + \abs{S_3}^2) - \underbrace{\frac{m_{12}^2}{3}
		\qty\Big(S_1^* S_2 + S_2^* S_3 + S_3^* S_1 + \text{h.c.})}_{\text{soft breaking}}
	\nonumber                                                                                       \\
	 & \quad +\frac{\lambda_S}{2} \qty\Big( \abs{S_1}^4 + \abs{S_2}^4 + \abs{S_3}^4 )
	+\lambda'_S \qty\Big(
		\abs{S_1}^2 \abs{S_2}^2
		+ \abs{S_2}^2 \abs{S_3}^2
		+ \abs{S_3}^2 \abs{S_1}^2
	)
	\nonumber                                                                                       \\
	 & \quad - \underbrace{\mu_\Phi^2 \abs{\Phi}^2
		+ \frac{\lambda_\Phi}{2} \abs{\Phi}^4}_\text{SM Higgs}
	+ \underbrace{\lambda_{\Phi S} \abs{\Phi}^2 \qty\Big(\abs{S_1}^2
			+ \abs{S_2}^2
			+ \abs{S_3}^2)}_{\text{Higgs portal}}.
\end{align}
We note that the potential has a dark CP symmetry $S_j \to S_j^*$.
The covariant derivatives are defined as
\begin{equation}\label{eq5}
	D_\mu S_j  = \qty(\partial_\mu - i g_V V_\mu)S_j \qquad (j=1,2,3),
\end{equation}
where $\Phi$ is the SM Higgs doublet, and $V_\mu$ is the gauge field associated with the dark $\mathrm{U}(1)_V$ symmetry. $V_{\mu\nu}$ and $Y_{\mu \nu}$ are the field strength tensor for the gauge bosons $V_\mu$ and $Y_\mu$ associated with the dark $\mathrm{U}(1)_V$ symmetry and SM $\mathrm{U}(1)_Y$, respectively.

\subsection{Residual symmetry in the broken phase}
\label{sec:residual-symmetry}

Without any loss of generality, one finds the following vacuum expectation values (VEVs) for the singlets\footnote{See appendix \ref{sec:VEV_analysis} for a detailed derivation of the VEVs.}
\begin{equation}
	\expval{S_1} = \expval{S_2} = \expval{S_3} = \frac{v_s}{\sqrt{6}},
\end{equation}
and for the Higgs doublet
\begin{equation}\label{eq2.8}
	\expval{\Phi} = \frac{1}{\sqrt{2}} \mqty(0 \\ v),
\end{equation}
where $v=1/\sqrt{\mathstrut \sqrt{2} G_F}\approx246~\mathrm{GeV}$ with the Fermi constant.
This configuration spontaneously breaks the local gauge $\mathrm{U}(1)_V$ and the global $\mathrm{U}(1)_A$ symmetries.

It is convenient to move to the so-called Higgs basis $\Sigma$ as
\begin{equation}\label{eq4}
	\Sigma = \mqty(\Sigma_1 \\ \Sigma_2 \\ \Sigma_3) =  R\cdot \mqty(S_1 \\ S_2\\ S_3),
\end{equation}
where $R$ is the rotation matrix of the $\mathbb{Z}_3$ symmetry \cite{Georgi:2000vve},
\begin{equation}
	R = \frac{1}{\sqrt{3}}\, \mqty(1 & 1 & 1 \\ 1 & \omega & \omega^2 \\ 1 & \omega^2 & \omega),
\end{equation}
with $\omega = e^{i 2 \pi/3}$.
In this basis, the following symmetries are manifested
\begin{align}
	\text{shift}\quad :         & \quad \Phi\to \Phi,\quad \Sigma \to T_1 \Sigma,\quad
	(\Sigma_1\to \Sigma_3,~\Sigma_2\to\Sigma_1,~\Sigma_3\to\Sigma_2),                  \\
	\text{clock~} \mathbb{Z}_3: & \quad \Phi\to \Phi,\quad \Sigma \to T_3 \Sigma,\quad
	(\Sigma_1\to \Sigma_1,~\Sigma_2\to\omega\Sigma_2,~\Sigma_3\to\omega^2\Sigma_3),
\end{align}
where a shift matrix $T_1$ and a clock matrix $T_3$ are introduced as
\begin{align}
	T_1=\mqty( 0 & 0 & 1 \\ 1 & 0 & 0 \\ 0            & 1 & 0 ),\quad
	T_3=\mqty( 1 & 0 & 0 \\ 0 & \omega & 0 \\ 0 & 0 & \omega^2 ),
\end{align}
with triality relations $T_1^3=T_3^3=1$. 
This decomposition shows that the $\Sigma_1$ is a $\mathbb{Z}_3$ singlet scalar while, $\Sigma_2$ and $\Sigma_3$ are charged under the $\mathbb{Z}_3$.
The $\mathbb{Z}_3$ symmetry remains unbroken in the vacuum configuration since $T_3\langle \Sigma \rangle = \langle \Sigma \rangle$.
Therefore, the $\mathbb{Z}_3$ symmetry ensures the stability of the DM candidate.
The dark scalar sector preserves dark CP symmetry even after spontaneous symmetry breaking, allowing for a simple analysis of its mass spectrum.

Substituting these VEVs into the scalar potential \cref{eq8} and minimizing the scalar potential, we obtain the following stationary conditions for $\mu_\Phi^2$ and $\mu_S^2$
\begin{align}
	\mu_\Phi^2
	 & = \frac{1}{2} v^2 \lambda_\Phi + \frac{1}{2} v_s^2 \lambda_{\Phi S}\label{eq14},                                     \\
	\mu_S^2
	 & = \frac{2}{3} m_{12}^2 - \frac{1}{6}	v_s^2 (\lambda_S + 2\lambda'_S) - \frac{1}{2} v^2 \lambda_{\Phi S}\label{eq15}.
\end{align}

\subsection{
	Mass spectrum}

The SM Higgs doublet fluctuation can be defined as
\begin{equation}
	\Phi = \frac{1}{\sqrt{2}} \mqty(0 \\ v + h(x)).
\end{equation}
Following the standard approach to separate the $\mathbb{Z}_3$ charges, we work in the Higgs basis as
\begin{equation}
	\mqty(\Sigma_1 \\ \Sigma_2 \\ \Sigma_3) \to
	R \cdot \qty{\frac{v_s}{\sqrt{6}}\mqty(1 \\ 1 \\ 1) + \mqty(S_1 \\ S_2 \\ S_3)}
	= \frac{v_s}{\sqrt{2}}\, \mqty(1 \\ 0 \\ 0) +
	\mqty(\Sigma_1 \\ \Sigma_2 \\ \Sigma_3).
\end{equation}
In this basis, only one of the complex scalar field gains VEV as
\begin{equation}
	\Sigma_1 = \frac{1}{\sqrt{2}} \qty(v_s + s'_1(x) + i z(x)),
\end{equation}
where $z$ is the would-be Nambu-Goldstone boson (NG) absorbed by the $\mathrm{U}(1)_V$ gauge boson. This allows a straightforward diagonalization of the mass matrix in the $\mathbb{Z}_3$ singlet sector, yielding the physical mass eigenstates $h_1$ and $h_2$ as
\begin{equation}
	\mqty(m_1^2 & 0 \\ 0 & m_2^2) =
	\mqty(
		c_\theta &  s_\theta\\
		-  s_\theta &  c_\theta
	)\mqty(
		v^2 \, \lambda_\Phi  &  v_s v\lambda_{\Phi S}\\
		v_s v\lambda_{\Phi S} & v_s^2 \qty(\lambda_S +2\lambda'_S)/3
	)\mqty(
		c_\theta & -  s_\theta\\
		s_\theta &  c_\theta
	),
\end{equation}
where $c_\theta = \cos\theta$, $s_\theta = \sin\theta$, and the mass eigenstate basis can be represented as
\begin{equation}
	\mqty( h_1 \\ h_2) =   \mqty(
		c_\theta &  s_\theta\\
		-  s_\theta &  c_\theta
	)\mqty(h \\ s'_1).
\end{equation}
Furthermore, the mixing angle $\theta$ for the mass eigenstates $h_1$ and $h_2$ is given by
\begin{equation}
	\tan 2\theta = \frac{2 v_s v\lambda_{\Phi S}}{v^2 \, \lambda_\Phi -  v_s^2 \flatfrac{\qty(\lambda_S +2\lambda'_S)}{3}}.
\end{equation}

On the other hand, the fields $\Sigma_2$ and $\Sigma_3^*$ are charged under the $\mathbb{Z}_3$ symmetry with a phase of $\omega$. Therefore, the mass matrix in terms of the mass eigenstates can be written as
\begin{equation}
	\mqty(m_{\Sigma}^2 & 0 \\ 0 & m_a^2) =
	\mqty(\frac{1}{\sqrt{2}} & \frac{1}{\sqrt{2}}\\-\frac{1}{\sqrt{2}} & \frac{1}{\sqrt{2}})
	\mqty(
		m_{12}^2 + \frac{v_s^2(\lambda_S -\lambda'_S)}{6} & \frac{v_s^2(\lambda_S -\lambda'_S)}{6}\\
		\frac{v_s^2(\lambda_S -\lambda'_S)}{6} & m_{12}^2 + \frac{v_s^2(\lambda_S -\lambda'_S)}{6}
	)
	\mqty(\frac{1}{\sqrt{2}} & - \frac{1}{\sqrt{2}}\\ \frac{1}{\sqrt{2}} & \frac{1}{\sqrt{2}}),
\end{equation}
where the mass eigenstate basis is determined as
\begin{align}\label{eq39}
	\mqty(\Sigma_\omega                 \\ a_\omega) & = \frac{1}{\sqrt{2}}
	\mqty(1 & 1                         \\-1 & 1) \mqty(\Sigma_2 \\ \Sigma_3^*),
	        &   & \mqty(\Sigma_\omega^* \\ a_\omega^*) = \frac{1}{\sqrt{2}}
	\mqty(1 & 1                         \\-1 & 1) \mqty(\Sigma_2^* \\ \Sigma_3),
\end{align}
since $\Sigma_\omega$ $(a_\omega)$ transforms as even (odd) under the dark CP parity.
Therefore, the masses are written as
\begin{align}
	m_{\Sigma}^2 & = m_{12}^2 + \frac{1}{3} v_s^2\qty\Big(\lambda_S -\lambda'_S),\label{eq2.41} \\
	m_a^2        & = m_{12}^2 \equiv m_{\text{DM}}^2.\label{eq2.42}
\end{align}

Thus, this setup yields one additional (dark) complex Higgs boson $\Sigma_\omega$ and a complex pNGB $a_\omega$. This $a_\omega$ field is naturally decoupled from the SM like Higgs field, making it a potential candidate for a DM particle that interacts weakly with visible matter.
Moreover, for the case of $\lambda_S =\lambda'_S$ we see that, these charged scalars are degenerate in mass due to the enhanced global symmetry, which might lead us to a multi-component DM candidate. We will discuss this in future work.

Therefore, the Lagrangian \cref{eq7} in mass eigenbasis will take the form:
\begin{equation}
	\mathcal{L} \supset -\frac{1}{2}m_1^2 h_1^2 -\frac{1}{2}m_2^2 h_2^2 - m_{\Sigma}^2 \abs{\Sigma_\omega}^2 - m_{\text{DM}}^2 \abs{a_\omega}^2.
\end{equation}

\subsection{Gauge Kinetic Mixing}

Following the same procedure as in \cite{Abe:2024vxz}, we get the mass eigenbasis for the gauge bosons as
\begin{align}
	\mqty(W^3_\mu                                            \\ Y'_\mu \\ V'_\mu)
	                     & = \mqty(
	c_\zeta^{}           & 0                    & s_\zeta^{} \\
	0                    & 1                    & 0          \\
	- s_\zeta^{}         & 0                    & c_\zeta^{}
	)\,\mqty(
	c_W^{}               & s_W^{}               & 0          \\
	- s_W^{}             & c_W^{}               & 0          \\
	0                    & 0                    & 1
	)\, \mqty(
	Z_\mu                                                    \\ A_\mu \\ Z'_\mu
	)
	\nonumber                                                \\
	                     & = \mqty(
	c_\zeta^{}  c_W^{}   & c_\zeta^{}  s_W^{}   & s_\zeta^{} \\
	- s_W^{}             & c_W^{}               & 0          \\
	- s_\zeta^{}  c_W^{} & - s_\zeta^{}  s_W^{} & c_\zeta^{}
	)\, \mqty(
	Z_\mu                                                    \\ A_\mu \\ Z'_\mu
	),
\end{align}
where $\zeta$ is given by
\begin{equation}
	\tan 2\zeta = \frac{2  s_\epsilon^{}  c_\epsilon^{}  s_W^{}}{\rho^2 - 1 +  s_\epsilon^2 \qty(1 +  s_W^2)},
\end{equation}
and $g_Z^{} = \sqrt{g^2 + g_Y^2},\, \rho^2 = \flatfrac{4 v_s^2 g^2_V}{v^2 g^2_Z},\, c_\zeta^{} = \cos\zeta,\, s_\zeta^{} = \sin\zeta,\, s_\epsilon^{} = \sin\epsilon,\, c_\epsilon^{} = \cos\epsilon,\, s_W^{} = \sin\theta_W^{},\, c_W^{} = \cos\theta_W^{}$ with the weak mixing angle $\theta_W^{}$.
In this basis, we can write the masses of the SM gauge boson $Z$ and dark gauge boson $Z'$ as
\begin{align}
	m^2_Z    & = \frac{v^2 g_Z^2}{4} \qty\Bigg{1 -  t_\epsilon  s_W  s_{2\zeta} +  s_\zeta^2 \qty(\frac{\rho^2}{ c_\epsilon^2} +  t_\epsilon^2  s_W^2 - 1)}, \\
	m^2_{Z'} & = \frac{v^2 g_Z^2}{4} \qty\Bigg{1 +  t_\epsilon  s_W  s_{2\zeta} +  c_\zeta^2 \qty(\frac{\rho^2}{ c_\epsilon^2} +  t_\epsilon^2  s_W^2 - 1)}.
\end{align}
From these expressions, we can find the dark gauge coupling as
\begin{equation}\label{eq-gV}
	g_V^{} = \frac{c_\epsilon}{2v_s}\sqrt{ 4\qty(m_Z^2 + m_{Z'}^2) - v^2 g_Z^2 \qty(1 + t_\epsilon^2 s_W^2) }.
\end{equation}

\subsection{Parameters of our model}

There are total nine parameters i.e., seven from the scalar sector and two from the new dark gauge sector in this model.
Apart from the stationary conditions in \cref{eq14} and \cref{eq15}, here we will write all the parameters of our model in terms of the physical mass eigenvalues, mixing angle and VEVs as
\begin{subequations}
	\begin{align}
		m_{12}^2 & = m_{\text{DM}}^2,                                                 \\
		\lambda_\Phi
		         & = \frac{m_1^2  c_\theta^2 + m_2^2  s_\theta^2}{v^2}\label{eq2.51}, \\
		\lambda_{\Phi S}
		         & = \frac{\qty(m_1^2 - m_2^2) s_\theta  c_\theta}{v_s v},            \\
		\lambda_S
		         & =  \frac{m_1^2  s_\theta^2 + m_2^2  c_\theta^2}{v_s^2}
		+ \frac{2\qty(m_\Sigma^2 - m_{\text{DM}}^2)}{v_s^2},                          \\
		\lambda'_S
		         & =   \frac{m_1^2  s_\theta^2 + m_2^2  c_\theta^2}{v_s^2}
		- \frac{m_\Sigma^2 - m_{\text{DM}}^2}{v_s^2}\label{eq2.54}.
	\end{align}
\end{subequations}
The remaining two parameters are the dark gauge coupling $g_V^{}$ (\ref{eq-gV}) and the gauge kinetic mixing angle $\epsilon$.
Thus, we are left with the following physical parameters
\begin{equation}
	m_1( = 125 \;\text{GeV}),\quad m_2,\quad \sin\theta,\quad m_{\text{DM}}^{},\quad
	m_\Sigma^{},\quad v, 
	\quad v_s, \quad m_{Z'}^{},	\quad \sin\epsilon.
	\label{eq2.30}
\end{equation}

\section{Constraints on the model}
\label{sec:constraints}

In this section, we discuss the constraints on the model parameters due to the Perturbative unitarity (PU) and the Higgs invisible decay.

\subsection{Perturbative unitarity}
\label{sec:constraints.1}
The PU of the model is a crucial aspect to ensure the stability of the theory. From the discussion of PU \cite{Lee:1977eg}, we can obtain the constraints on the model parameters. Since we are going to deal with high energy scattering processes, we study the Lagrangian in the symmetric phase \cref{eq8}. For this purpose, we define the SM Higgs doublet as $\Phi^T = \mqty(H_1 & H_2)$. Therefore, we can construct the charge-neutral states as
\begin{equation}
	i\to f \quad \forall\;\; i,f \in \qty{H_1 H_1^*,\, H_2 H_2^*,\, S_1 S_1^*,\, S_2 S_2^*,\, S_3 S_3^*,\, S_1 S_2^*,\, S_2 S_3^*,\, S_3 S_1^* }.
\end{equation}

The partial wave matrix $a^0$ thus can be constructed as
\begin{align}
	(a^0)_{fi} = \frac{1}{16 \pi} \mqty(
	2 \lambda_\Phi   & \lambda_\Phi     & \lambda_{\Phi S} & \lambda_{\Phi S} & \lambda_{\Phi S} & 0          & 0          & 0          \\
	\lambda_\Phi     & 2\lambda_\Phi    & \lambda_{\Phi S} & \lambda_{\Phi S} & \lambda_{\Phi S} & 0          & 0          & 0          \\
	\lambda_{\Phi S} & \lambda_{\Phi S} & 2\lambda_S       & \lambda'_S       & \lambda'_S       & 0          & 0          & 0          \\
	\lambda_{\Phi S} & \lambda_{\Phi S} & \lambda'_S       & 2\lambda_S       & \lambda'_S       & 0          & 0          & 0          \\
	\lambda_{\Phi S} & \lambda_{\Phi S} & \lambda'_S       & \lambda'_S       & \lambda_S        & 0          & 0          & 0          \\
	0                & 0                & 0                & 0                & 0                & \lambda'_S & 0          & 0          \\
	0                & 0                & 0                & 0                & 0                & 0          & \lambda'_S & 0          \\
	0                & 0                & 0                & 0                & 0                & 0          & 0          & \lambda'_S \\
	).
\end{align}
Calculating the eigenvalues of this matrix and imposing the PU conditions, we find the following inequalities
\begin{subequations}
	\begin{align}
		\abs{\lambda_\Phi}             & < 8\pi,   \\
		\abs{\lambda_S' - 2 \lambda_S} & < 8 \pi,  \\
		\abs{\lambda_S'}               & < 8 \pi,  \\
		\abs{3 \lambda_\Phi  + 2 \lambda_S'
			+ 2 \lambda_S
			\pm \sqrt{24 \lambda_{\Phi S}^2 +
				\qty(-3 \lambda_\Phi
					+ 2 \lambda_S'
					+ 2 \lambda_S)^2}
		}                              & < 16 \pi.
	\end{align}
\end{subequations}
From the charged states under the $\mathrm{U}(1)_V$ symmetry, we have the channels such as $S_1 S_1 \leftrightarrow S_1 S_1$ and $S_1 H_1 \leftrightarrow S_1 H_1$, which will give us the following inequalities:
\begin{subequations}
	\begin{align}
		\abs{\lambda_S}        & < 8 \pi, \\
		\abs{\lambda_{\Phi S}} & < 8 \pi.
	\end{align}
\end{subequations}

The PU bounds on the gauge couplings are similar to the ones in \cite{Abe:2024vxz}
\begin{equation}
	g_V < \sqrt{4 \pi}.
\end{equation}

\subsection{Higgs invisible decay}

\begin{figure}[t]
	\centering
	\begin{equation*}
		\raisebox{-0.45\height}{\includegraphics[width=2cm]{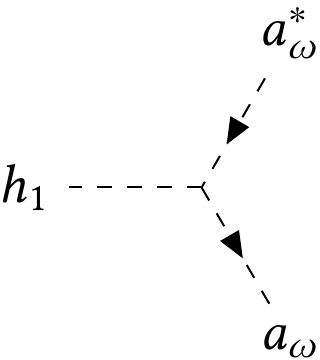}}
		= i \kappa_{aah_1}
	\end{equation*}

	\caption{SM Higgs--DM--DM vertex}
	\label{fig:aah vertex}
\end{figure}

A stringent constraint on the parameter space arises from the Higgs invisible decay width, particularly when $m_{\text{DM}} < m_1/2$. In this case, the Higgs boson $h_1$ can decay into a pair of DM particles $a_\omega a_\omega^*$, contributing to the invisible decay channel. The corresponding vertex is as shown in \cref{fig:aah vertex}, where
\begin{equation}
	\kappa_{aah_1} = v \lambda_{\Phi S} \cos \theta  + \frac{1}{3}v_s \qty(\lambda_S + 2 \lambda_S') \sin \theta.
\end{equation}
Thus the decay width is calculated as
\begin{equation}
	\Gamma(h_1 \to a_\omega a_\omega) = \frac{1}{32 \pi} \frac{\kappa_{aah_1}^2}{m_1} \sqrt{1 - \frac{4 m_{\text{DM}}^2}{m_1^2}} \; \Theta\qty(m_1 - 2 m_{\text{DM}}^{}).
\end{equation}
Here, $\Theta\qty(m_1 - 2 m_{\text{DM}})$ is the Heaviside step function, which ensures that the decay only occurs when $m_1 > 2 m_{\text{DM}}$.

This process is being searched by ATLAS \cite{ATLAS:2023tkt} and CMS \cite{CMS:2023sdw} experiments with the upper bound currently at,
\begin{equation}
	\text{BR}_{\text{inv}} <
	\begin{cases}
		0.107 & (\text{ATLAS}) \\
		0.15  & (\text{CMS})
	\end{cases}.
\end{equation}







\section{Dark matter}
\label{sec:DM}

\subsection{New channels}
\label{sec:new_channels}

\begin{figure}[t]
	\centering
	\begin{subfigure}{0.45\textwidth}
		\centering
		\includegraphics[width=0.48\textwidth]{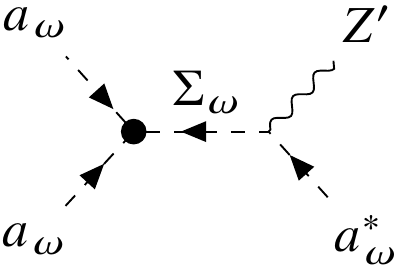}
		\hfill
		\includegraphics[width=0.32\textwidth]{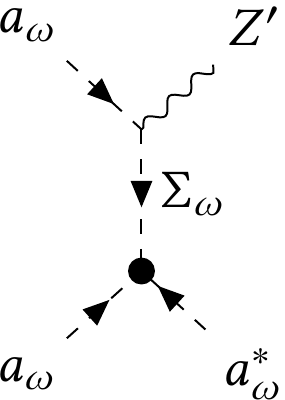}
		\caption{semi-annihilation channel}
	\end{subfigure}
	\hfill
	\begin{subfigure}{0.45\textwidth}
		\centering
		\includegraphics[width=0.48\textwidth]{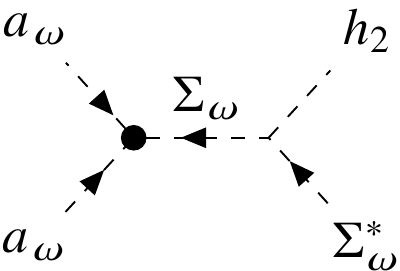}
		\hfill
		\includegraphics[width=0.32\textwidth]{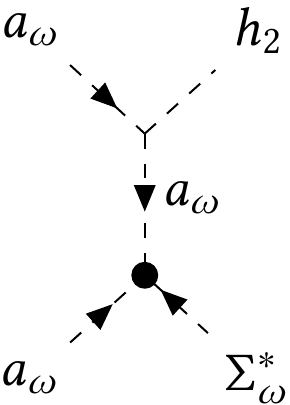}
		\caption{semi-annihilation-like channel}
	\end{subfigure}
	\caption{Semi-annihilation processes arising from the new non-trivial cubic interaction terms in the scalar sector.}
	\label{fig:semi-annihilation}
\end{figure}

In $\mathbb{Z}_3$ symmetric DM models, semi-annihilation processes \cite{Hambye:2008bq, DEramo:2010keq}
are possible due to the presence of the cubic terms in the scalar potential, which are forbidden in the $\mathbb{Z}_2$ models. The Lagrangian governing these processes can be expressed as
\begin{align}
	\mathcal{L}  \supset
	\kappa_1
	\qty\Big(\Sigma_\omega \overset{\leftrightarrow}{\partial_\mu} a_\omega^*
		+ a_\omega \overset{\leftrightarrow}{\partial_\mu} \Sigma_\omega^*) Z'^\mu
	-
	\kappa_2
	\underbrace{\qty\Big(\Sigma_\omega^3 + \Sigma_\omega^{*3} - \Sigma_\omega a_\omega^2 - \Sigma_\omega^* a_\omega^{*2})}_\text{Unique cubic interactions}
	-\;
	\kappa_3
	\abs{\Sigma_\omega}^2  \; h_2,
\end{align}
where $A\overset{\leftrightarrow}{\partial_\mu}B=A\partial_\mu B-B\partial_\mu A$, and
\begin{equation}
	\kappa_1 = \frac{g_V^{} c_\zeta}{c_\epsilon}, \qquad
	\kappa_2 = \frac{m_\Sigma^2 - m_{\text{DM}}^2}{2 v_s}, \qquad
	\kappa_3 = \frac{m_2^2 + 2\qty(m_\Sigma^2 - m_{\text{DM}}^{2})}{v_s}  \cos\theta.
\end{equation}
Among these, the $\kappa_2$ term encapsulates the genuinely new cubic interactions that are unique to our model, which leads to the following semi-annihilation processes
\begin{equation}
	\label{eq:semi-annihilation}
	a_\omega a_\omega \to \Sigma_\omega \to a_\omega^* Z', \qquad
	a_\omega a_\omega \to \Sigma_\omega \to \Sigma_\omega^* h_2. \qquad
\end{equation}
The Feynman diagrams for these processes are shown in \cref{fig:semi-annihilation}.

In contrast to the $\mathbb{Z}_2$ stabilized DM, where $\text{DM} +\text{DM} \to \text{SM} + \text{SM}$ annihilation processes are dominant, $\mathbb{Z}_3$ models open up new semi-annihilation channels $\text{DM} +\text{DM} \to \text{DM} + \text{X}$ where $\text{X}$ is either a SM particle or a new particle. These offer alternative mechanisms for thermal freeze-out and can lead to qualitatively distinct indirect detection signals, thus enabling viable relic density even where standard annihilation channels are inefficient.
This semi-annihilation structure also makes the model a natural candidate for exploring BDM signals.

\subsection{Relic abundance}
\label{sec:relic}

The DM relic abundance in our model is determined by the thermal freeze-out mechanism. The complex scalar $a_\omega$, stabilized by a residual $\mathbb{Z}_3$ symmetry, was once in thermal equilibrium with the SM plasma. As the universe expanded and cooled, the interaction rate of $a_\omega$ dropped below the Hubble rate, causing it to decouple and freeze out with a relic density set by its annihilation and semi-annihilation cross sections.

In addition, it is worth mentioning that the direct detection signals are naturally suppressed in this model.
The tree-level scattering amplitude between the DM and nucleons exhibits a cancellation between the SM-like Higgs $h_1$ and the additional scalar $h_2$ exchanges at low momentum transfer, a characteristic feature of pNGB DM models \cite{Gross:2017dan}. Therefore, the direct detection constraints are automatically satisfied without requiring fine-tuning of parameters.

Similar to the $\mathbb{Z}_2$ model \cite{Abe:2024vxz}, the annihilation processes are mediated by the $s$-channel exchange of the SM-like Higgs boson $h_1$ and the additional scalar $h_2$. And considering a light dark gauge boson $m_{Z'}^{} = \order{100}\,\text{GeV}$, the $a_\omega a_\omega^* \to Z'Z'$ annihilation channel will open, which will give us a large annihilation cross section. But in this model due to the presence of the semi-annihilation process \cref{eq:semi-annihilation} i.e., $a_\omega a_\omega \to a_\omega^* Z'$, the DM candidate $a_\omega$ can also semi-annihilate into a $Z'$ boson and a BDM, which increases the overall cross-section when $m_{\text{DM}}^{} > m_{Z'}^{}$.

\begin{figure}[t]
	\centering
	\includegraphics[width=0.8\textwidth]{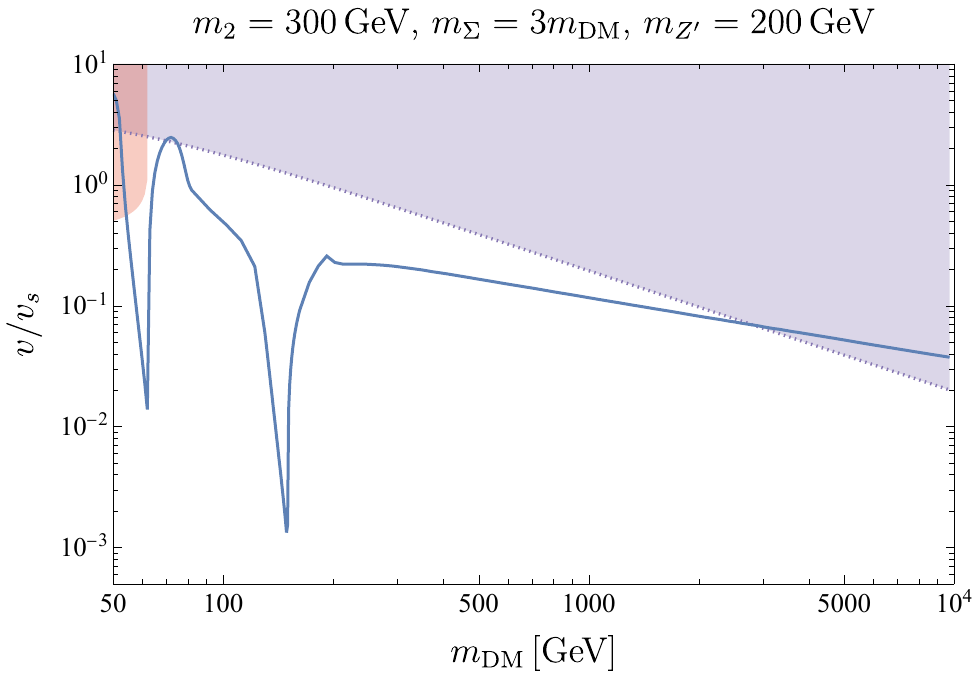}
	\caption{Prediction of viable parameter space to explain the observed DM relic abundance. We choose $\sin \theta = 0.1,\, \sin \epsilon = 10^{-4},\, m_2 = 300\,\text{GeV},\, m_\Sigma^{} = 3 m_{\text{DM}}^{}$, and $m_{Z'}^{} = 200\,\text{GeV}$. The thick blue line represents the correct relic abundance $\Omega_{\text{DM}} h^2 = 0.12 \pm 0.001$. The red region is excluded by the Higgs invisible decay and the blue region is excluded by the PU bounds for the scalar quartic and dark gauge coupling.}
	\label{fig:relic abundance}
\end{figure}

We choose the parameters in \cref{eq2.30} to be $\sin \theta = 0.1,\;$ $\sin \epsilon = 10^{-4},\;$ $m_2 = 300\,\text{GeV},\;$ $m_\Sigma^{} = 3 m_{\text{DM}}^{}$, and $m_{Z'}^{} = 200\,\text{GeV}$.
We use \texttt{FeynRules} \cite{Alloul:2013bka} to generate model files and \texttt{micrOMEGAs} \cite{Alguero:2023zol} to calculate the DM relic abundance.
As expected, the behavior of the relic abundance shown in \cref{fig:relic abundance} closely resembles that of the $\mathbb{Z}_2$ model when $m_{\Sigma}^{} = 3\,m_{\text{DM}}^{}$.
The thick blue line represents the relic abundance in $m_{\text{DM}}^{}\,\text{vs.}\, (\flatfrac{v}{v_s})$ and is consistent with the observed value of $\Omega_{\text{DM}} h^2 = 0.12 \pm 0.001$ \cite{Planck:2018vyg}.
The red region is excluded by the Higgs invisible decay and the blue region is excluded by the PU bounds for the scalar quartic and dark gauge coupling (see \cref{sec:constraints.1}).
In our case, the PU bounds are significant and restrict the model up to $m_{\text{DM}}^{} \lesssim 2.5\,\text{TeV}$ because of the dependence of the scalar quartic couplings.
Moreover, as usual, there are two dips in the relic abundance curve, which corresponds to the $h_1$ and $h_2$ $s$-channel resonances $m_{\text{DM}}^{} = \flatfrac{125}{2}\,\text{GeV}$ and
$\flatfrac{300}{2}\,\text{GeV}$, respectively.
The visible kink at $m_{\text{DM}}^{} = m_{Z'}^{} = 200\,\text{GeV}$ reflects the kinematic threshold for the $Z'$ production, where the $a_\omega a_\omega^* \to Z'Z'$ annihilation channel opens up.

\begin{figure}[t]
	\centering
	\includegraphics[width=0.8\textwidth]{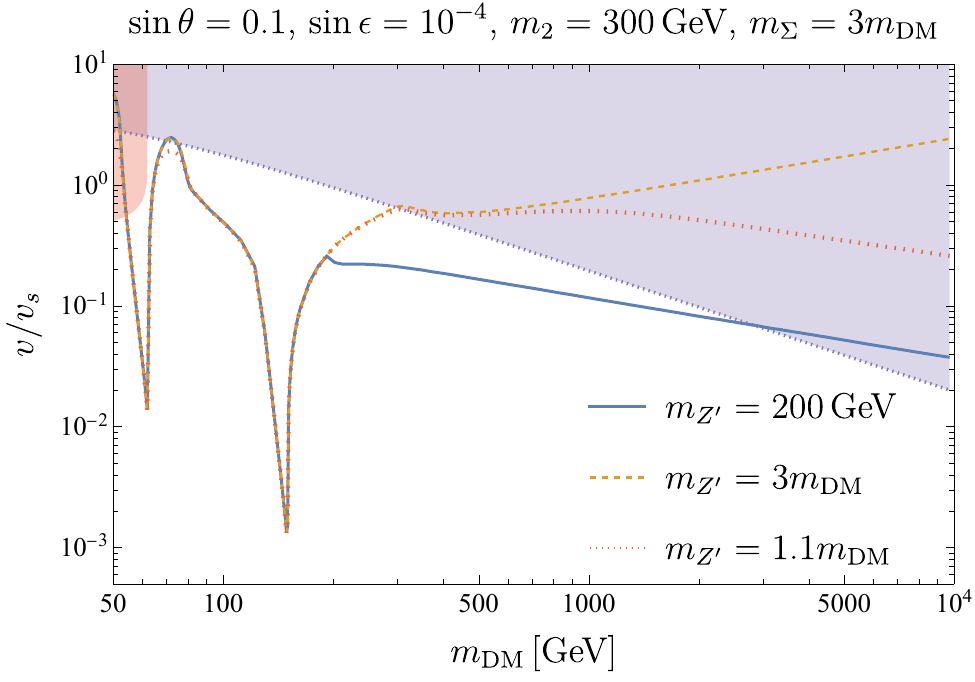}
	\caption{Comparison of relic abundance predictions for different values of the dark gauge boson mass $m_{Z'}^{}$. The red region is excluded by the Higgs invisible decay and the blue region is excluded by the PU bounds due to the dependence on the scalar quartic couplings in all three cases. The thick blue curve corresponds to $m_{Z'}^{} = 200\,\text{GeV}$, while the dashed yellow curve shows the case for $m_{Z'}^{} = 3\,m_{\text{DM}}^{}$. Semi-annihilation effects and kinematically allowed $Z'$ channels significantly affect the relic density across the parameter space.}
	\label{fig:relic abundance comparison}
\end{figure}

We also discuss the behavior of the relic abundance with varying $Z'$ mass. In \cref{fig:relic abundance comparison},
the thick blue line corresponds to the same curve in \cref{fig:relic abundance} and is compared with the dashed orange curve for $m_{Z'}^{}  = 3 m_{\text{DM}}^{}$.
In the case of near degeneracy i.e., $m_{Z'}^{} = 1.1 m_{\text{DM}}^{}$, the forbidden $Z'Z'$ channel remains open, therefore leading to a large cross section.
In addition, the semi-annihilation process $a_\omega a_\omega \to a_\omega^* Z'$ further enhances the cross section,
resulting in the sharp decrease in the relic abundance curve compared to the $\mathbb{Z}_2$ model.

\begin{figure}[t]
	\centering
	\includegraphics[width=0.8\textwidth]{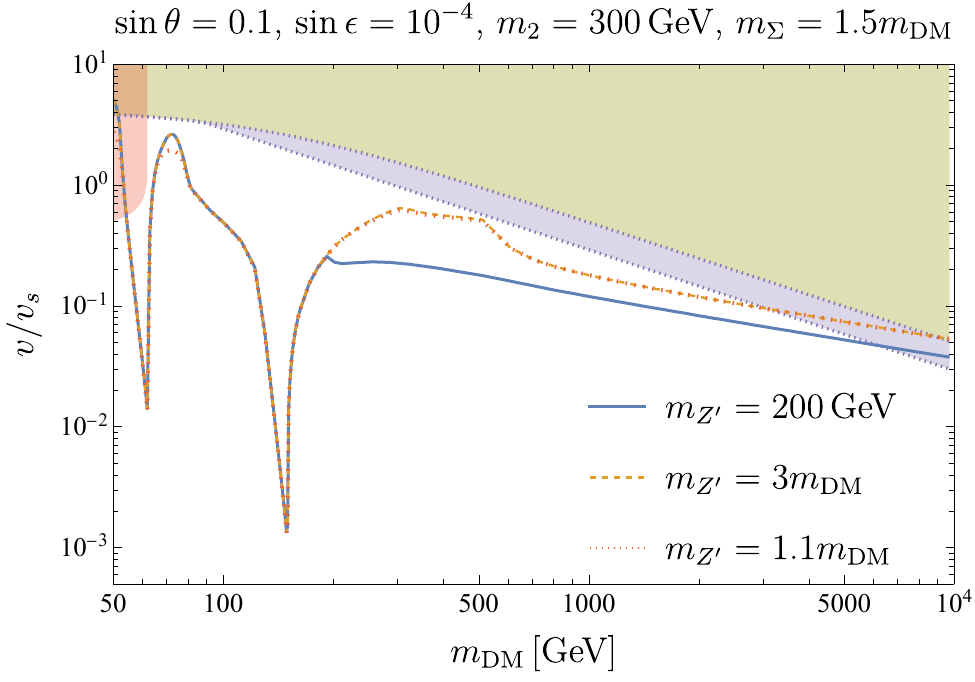}
	\caption{Comparison of relic abundance predictions assuming $m_\Sigma^{} = 1.5\,m_{\text{DM}}^{}$, mimicking a $\mathbb{Z}_2$-like spectrum. The red region is excluded by the Higgs invisible decay, while the blue region is excluded by the PU bounds on the scalar quartic couplings for $m_{Z'} = 200\,\text{GeV}$ and the green region corresponds to the PU bounds for the other two cases where the dark gauge coupling also contributes.}
	\label{fig:relic abundance sigma}
\end{figure}

If we try to recreate the same scenario in $\mathbb{Z}_2$ model by choosing $m_\Sigma^{} = 1.5 m_{\text{DM}}^{}$ then we get the curve as shown in \cref{fig:relic abundance sigma}.
Here, except for the $m_{Z'}^{} = 200\,\text{GeV}$ case, a new forbidden semi-annihilation-like process $a_\omega a_\omega \to \Sigma_\omega^* h_2$ (see \cref{eq:semi-annihilation}) becomes active in our model,
leading to a sudden enhancement of the cross section compared to the $\mathbb{Z}_2$ model.
This highlights that the presence of such semi-annihilation-like processes results in qualitatively distinct relic abundance behaviors.


\subsection{Phenomenological impacts}
The first process of \cref{eq:semi-annihilation} ($a_\omega a_\omega \to a_\omega^* Z^\prime$) could, in principle, lead to a flux of relativistic $a_\omega$ particles
that scatter in large-volume neutrino detectors~\cite{DUNE:2020ypp} and DM direct detection experiments~\cite{XENON:2023cxc},
as in refs.~\cite{Aoki:2023tlb, BetancourtKamenetskaia:2025noa}.
Assuming the DM particles in the initial state are non-relativistic, the energy of the DM particle in the final state is kinematically determined as
\begin{align}
	E_\text{DM}^{}=\frac{5}{4}m_\text{DM}^{}\left(1-\frac{m_{Z^\prime}^2}{5m_\text{DM}^2}\right).
\end{align}
The differential cross section for the elastic scattering with a nucleon ($a_\omega N\to a_\omega N$) can be calculated as
\begin{align}
	\frac{d\sigma_\text{el}}{dt}=\frac{f_N^2\sin^22\theta\, m_N^2}{32\pi v^2v_s^2\lambda(s,m_\text{DM}^2,m_N^2)}\frac{t^2\left(m_1^2-m_2^2\right)^2\left(4m_N^2-t\right)}{\left(t-m_1^2\right)^2\left(t-m_2^2\right)^2},
	\label{eq:dsdt}
\end{align}
where $s$ and $t$ are the Mandelstam variables, $m_N^{}=0.938~\mathrm{GeV}$ is the nucleon mass, $f_N^{}\sim0.3$ is the Higgs--nucleon coupling and
$\lambda(x,y,z)=x^2+y^2+z^2-2xy-2yz-2zx$ is the kinematic function.
The total cross section can be obtained by integrating eq.~(\ref{eq:dsdt}) over $t$.
Since $m_N^{}\ll m_\text{DM}^{}$, it is simplified and is given by
\begin{align}
	\sigma_\text{el}^{}\approx
	\frac{f_N^2\sin^22\theta\, m_N^4}{24\pi v^2v_s^2 s^3}\frac{\left(m_1^2-m_2^2\right)^2}{m_1^4m_2^4}\left(s-m_\text{DM}^2-m_N^2\right)^4v_\text{DM}^4,
\end{align}
where $v_\text{DM}^{}=\sqrt{1-m_\text{DM}^2/E_\text{DM}^2}\leq 3/5$ is the DM speed, and $s=m_\text{DM}^2+m_N^2+2E_\text{DM}^{}m_N^{}$.
Taking a typical set of the parameters i.e., $\sin\theta=0.1$, $v_s= v$, $m_2=300~\mathrm{GeV}$, $m_\text{DM}^{}=100~\mathrm{GeV}$, and $m_{Z^\prime}^{}\ll m_\text{DM}^{}$,
one can find the magnitude of the total cross section as $\sigma_\text{el}\sim 10^{-54}\,\text{cm}^2$,
which is several orders of magnitude below the experimental sensitivity \cite{DUNE:2020fgq,LZ:2022lsv,XENON:2023cxc}.
As a result, the detection rate is negligibly small, and no observable signal is expected in terrestrial detectors within this minimal setup.
In fact, the smallness of the cross section is due to the moderate boost of DM ($\gamma_\mathrm{DM}^{} \equiv \flatfrac{E_\mathrm{DM}^{}}{m_\mathrm{DM}^{}} \leq \flatfrac{5}{4}$).

Once the DM particles are highly boosted somehow, for example by decays or annihilations of the particles much heavier than the DM, one can expect that the cross section is highly enhanced, and detectable signals could be obtained. For example, if the boost factor $\gamma_\mathrm{DM}=30$ is achieved, we can expect the cross section to be of the order of $\sigma_\mathrm{el}\sim10^{-45}~\mathrm{cm}^2$, which can reach the current and future experimental sensitivity. Note in that case, deep inelastic scatterings will be dominant processes, but not the above elastic scattering. Therefore further modification with the hadronic part of the matrix element is required for precise calculations \cite{Paschos:2001np,Berger:2019ttc,HoefkenZink:2024hor}.

The Higgs bosons $h_i$ and the $Z^\prime$ gauge boson are also produced via semi-annihilation and semi-annihilation like processes in \cref{fig:semi-annihilation}, and they subsequently decay into SM particles.
These decay products can be detected through cosmic-ray and gamma-ray observations, similarly to those from standard dark matter annihilation.
The strength of indirect detection bounds depends on the annihilation channels and the dark matter halo profile.
For the standard annihilation into $\tau\overline{\tau}$ and $q\overline{q}$ final states, the dark matter mass region below 100 GeV is constrained \cite{Fermi-LAT:2015att}.
In our case, since gamma rays are produced via cascade decays, their energy spectrum is softer, and consequently, the indirect detection bounds are expected to be weaker.

In addition, semi-annihilation processes may induce dark matter self-heating, which could alleviate small-scale structure problems such as the core-cusp, missing satellite, and too-big-to-fail problems, particularly if the dark matter mass lies in the sub-GeV range. However, in our scenario, the dark matter mass is assumed to be above the electroweak scale in connection with direct detection experiments \cite{Chu:2018nki}. Nonetheless, such self-heating effects may still impact on precise calculations of the dark matter relic abundance \cite{Kamada:2017gfc}.

\section{Conclusion}
\label{sec:conclusion}

In this work, we have proposed a pseudo-Nambu-Goldstone boson dark matter model stabilized by a residual $\mathbb{Z}_3$ symmetry, realized through three complex scalar fields charged under a dark $\mathrm{U}(1)_V$ symmetry. The model naturally suppresses direct detection signals via symmetry-induced cancellations and introduces semi-annihilation channels not accessible in $\mathbb{Z}_2$ scenarios.

We examined theoretical and experimental constraints, including perturbative unitarity and Higgs invisible decay bounds, and confirmed that a viable parameter space exists. The relic abundance was computed using \texttt{micrOMEGAs}, showing that both annihilation and semi-annihilation processes significantly contribute to determine the relic abundance.

While the model was partly motivated by the possibility of generating boosted dark matter signals detectable at neutrino detectors, we found the resulting cross section to be well below the current experimental sensitivities. Nevertheless, this setup provides a minimal and consistent framework for semi-annihilation-driven pNGB BDM and opens the path for more predictive extensions.

\acknowledgments

This work was supported in part by JSPS Grant-in-Aid for Scientific Research KAKENHI Grants
No. JP22K03620 (K.T.), 25H02179 (T.T.) and 25K07279 (T.T.).

\appendix

\section{VEV analysis}
\label{sec:VEV_analysis}

We parameterize the most general VEVs for the singlets as
\begin{equation}\label{eq2.6}
	\expval{S_1} = \frac{v_{s_1}}{\sqrt{2}}, \quad
	\expval{S_2} = \frac{v_{s_2}}{\sqrt{2}}, \quad
	\expval{S_3} = \frac{v_{s_3}}{\sqrt{2}},
\end{equation}
and the VEV for the Higgs doublet is as shown in \cref{eq2.8}. This configuration spontaneously breaks the local gauge $\mathrm{U}(1)_V$ symmetry. Substituting these VEVs into the scalar potential of \cref{eq8} and minimizing it, we obtain the following equations
\begin{align}
	\pdv{\expval{\mathcal{V}}}{v}
	 & =
	\qty{-\mu_\Phi^2 + \frac{\lambda_\Phi}{2} v^2 + \frac{\lambda_{\Phi S}}{2} \qty(v_{s_1}^2 + v_{s_2}^2 + v_{s_3}^2)} v = 0,
	\\
	\pdv{\expval{\mathcal{V}}}{v_{s_1}}
	 & =
	\qty{\mu_S^2 + \frac{\lambda_{S}}{2} v_{s_1}^2 + \frac{\lambda'_S}{2} \qty( v_{s_2}^2 + v_{s_3}^2) + \frac{\lambda_{\Phi S}}{2} v^2 - \frac{m_{12}^2}{3}\qty(\frac{v_{s_2} + v_{s_3}}{v_{s_1}})}v_{s_1} = 0,
	\\
	\pdv{\expval{\mathcal{V}}}{v_{s_2}}
	 & =
	\qty{\mu_S^2 + \frac{\lambda_{S}}{2} v_{s_2}^2  + \frac{\lambda'_S}{2} \qty(v_{s_3}^2 + v_{s_1}^2) + \frac{\lambda_{\Phi S}}{2} v^2 - \frac{m_{12}^2}{3}\qty(\frac{v_{s_3} + v_{s_1}}{v_{s_2}})}v_{s_2} = 0,
	\\
	\pdv{\expval{\mathcal{V}}}{v_{s_3}}
	 & =
	\qty{\mu_S^2 + \frac{\lambda_{S}}{2}  v_{s_3}^2 + \frac{\lambda'_S}{2} \qty(v_{s_1}^2 + v_{s_2}^2) + \frac{\lambda_{\Phi S}}{2} v^2 - \frac{m_{12}^2}{3}\qty(\frac{v_{s_1} + v_{s_2}}{v_{s_3}})}v_{s_3} = 0.
\end{align}
To break the dark symmetry, at least one of the VEVs must be nonzero (e.g., $v_{s_1} \ne 0$), leading to two nontrivial minima for the VEVs as follows:

\paragraph{\boldmath $1.\, v_{s_1} \ne 0,\,  v_{s_2} = 0,\, v_{s_3} = - v_{s_1}$.}

When one of the singlet VEVs is zero, in addition to a non-zero VEV, the third becomes dependent on the non-zero VEV. Such a configuration, even though satisfies all four minimization equations, does not preserve the residual $\mathbb{Z}_3$ symmetry, as it breaks the permutation symmetry among the singlet VEVs. Therefore, we do not consider this vacuum further.

\paragraph{\boldmath $2. \, v_{s_1} = v_{s_2} = v_{s_3} \ne  0$.}
When none of the VEVs vanishes, it follows that
\begin{equation}
	v_{s_1} = v_{s_2} = v_{s_3} = \frac{v_s}{\sqrt{3}},
\end{equation}
which leads to the stationary conditions as shown in \cref{eq14,eq15}. This demonstrates that the vacuum configuration with equal VEVs for all the singlets preserves the permutation symmetry, as required by our construction. Such an alignment of VEVs are consistent with the residual $\mathbb{Z}_3$ symmetry, ensuring the stability of the DM candidate as discussed in \cref{sec:residual-symmetry}.

\bibliographystyle{JHEP}
\bibliography{reference.bib}

\providecommand{\href}[2]{#2}\begingroup\raggedright\begin{thebibliography}{10}

\bibitem{PandaX-4T:2021bab}
{\scshape PandaX-4T} collaboration, \emph{{Dark Matter Search Results from the PandaX-4T Commissioning Run}}, \href{https://doi.org/10.1103/PhysRevLett.127.261802}{\emph{Phys. Rev. Lett.} {\bfseries 127} (2021) 261802} [\href{https://arxiv.org/abs/2107.13438}{{\ttfamily 2107.13438}}].

\bibitem{LZ:2022lsv}
{\scshape LZ} collaboration, \emph{{First Dark Matter Search Results from the LUX-ZEPLIN (LZ) Experiment}}, \href{https://doi.org/10.1103/PhysRevLett.131.041002}{\emph{Phys. Rev. Lett.} {\bfseries 131} (2023) 041002} [\href{https://arxiv.org/abs/2207.03764}{{\ttfamily 2207.03764}}].

\bibitem{XENON:2023cxc}
{\scshape XENON} collaboration, \emph{{First Dark Matter Search with Nuclear Recoils from the XENONnT Experiment}}, \href{https://doi.org/10.1103/PhysRevLett.131.041003}{\emph{Phys. Rev. Lett.} {\bfseries 131} (2023) 041003} [\href{https://arxiv.org/abs/2303.14729}{{\ttfamily 2303.14729}}].

\bibitem{Gross:2017dan}
C.~Gross, O.~Lebedev and T.~Toma, \emph{{Cancellation Mechanism for Dark-Matter{\textendash}Nucleon Interaction}}, \href{https://doi.org/10.1103/PhysRevLett.119.191801}{\emph{Phys. Rev. Lett.} {\bfseries 119} (2017) 191801} [\href{https://arxiv.org/abs/1708.02253}{{\ttfamily 1708.02253}}].

\bibitem{Planck:2018vyg}
{\scshape Planck} collaboration, \emph{{Planck 2018 results. VI. Cosmological parameters}}, \href{https://doi.org/10.1051/0004-6361/201833910}{\emph{Astron. Astrophys.} {\bfseries 641} (2020) A6} [\href{https://arxiv.org/abs/1807.06209}{{\ttfamily 1807.06209}}].

\bibitem{Karamitros:2019ewv}
D.~Karamitros, \emph{{Pseudo Nambu-Goldstone Dark Matter: Examples of Vanishing Direct Detection Cross Section}}, \href{https://doi.org/10.1103/PhysRevD.99.095036}{\emph{Phys. Rev. D} {\bfseries 99} (2019) 095036} [\href{https://arxiv.org/abs/1901.09751}{{\ttfamily 1901.09751}}].

\bibitem{Abe:2020iph}
Y.~Abe, T.~Toma and K.~Tsumura, \emph{{Pseudo-Nambu-Goldstone dark matter from gauged $U(1)_{B-L}$ symmetry}}, \href{https://doi.org/10.1007/JHEP05(2020)057}{\emph{JHEP} {\bfseries 05} (2020) 057} [\href{https://arxiv.org/abs/2001.03954}{{\ttfamily 2001.03954}}].

\bibitem{Okada:2020zxo}
N.~Okada, D.~Raut and Q.~Shafi, \emph{{Pseudo-Goldstone dark matter in a gauged $B-L$ extended standard model}}, \href{https://doi.org/10.1103/PhysRevD.103.055024}{\emph{Phys. Rev. D} {\bfseries 103} (2021) 055024} [\href{https://arxiv.org/abs/2001.05910}{{\ttfamily 2001.05910}}].

\bibitem{Abe:2021byq}
Y.~Abe, T.~Toma, K.~Tsumura and N.~Yamatsu, \emph{{Pseudo-Nambu-Goldstone dark matter model inspired by grand unification}}, \href{https://doi.org/10.1103/PhysRevD.104.035011}{\emph{Phys. Rev. D} {\bfseries 104} (2021) 035011} [\href{https://arxiv.org/abs/2104.13523}{{\ttfamily 2104.13523}}].

\bibitem{Okada:2021qmi}
N.~Okada, D.~Raut, Q.~Shafi and A.~Thapa, \emph{{Pseudo-Goldstone dark matter in SO(10)}}, \href{https://doi.org/10.1103/PhysRevD.104.095002}{\emph{Phys. Rev. D} {\bfseries 104} (2021) 095002} [\href{https://arxiv.org/abs/2105.03419}{{\ttfamily 2105.03419}}].

\bibitem{Abe:2022mlc}
T.~Abe and Y.~Hamada, \emph{{A model of pseudo-Nambu{\textendash}Goldstone dark matter from a softly broken SU(2) global symmetry with a U(1) gauge symmetry}}, \href{https://doi.org/10.1093/ptep/ptad021}{\emph{PTEP} {\bfseries 2023} (2023) 033B04} [\href{https://arxiv.org/abs/2205.11919}{{\ttfamily 2205.11919}}].

\bibitem{Liu:2022evb}
D.-Y.~Liu, C.~Cai, X.-M.~Jiang, Z.-H.~Yu and H.-H.~Zhang, \emph{{Ultraviolet completion of pseudo-Nambu-Goldstone dark matter with a hidden U(1) gauge symmetry}}, \href{https://doi.org/10.1007/JHEP02(2023)104}{\emph{JHEP} {\bfseries 02} (2023) 104} [\href{https://arxiv.org/abs/2208.06653}{{\ttfamily 2208.06653}}].

\bibitem{Otsuka:2022zdy}
H.~Otsuka, T.~Shimomura, K.~Tsumura, Y.~Uchida and N.~Yamatsu, \emph{{Pseudo-Nambu-Goldstone dark matter from non-Abelian gauge symmetry}}, \href{https://doi.org/10.1103/PhysRevD.106.115033}{\emph{Phys. Rev. D} {\bfseries 106} (2022) 115033} [\href{https://arxiv.org/abs/2210.08696}{{\ttfamily 2210.08696}}].

\bibitem{Abe:2024vxz}
T.~Abe, Y.~Hamada and K.~Tsumura, \emph{{A model of pseudo-Nambu-Goldstone dark matter with two complex scalars}}, \href{https://doi.org/10.1007/JHEP05(2024)076}{\emph{JHEP} {\bfseries 05} (2024) 076} [\href{https://arxiv.org/abs/2401.02397}{{\ttfamily 2401.02397}}].

\bibitem{Abe:2024lzj}
T.~Abe and K.~Ichiki, \emph{{Tiny yet detectable WIMP-nucleon scattering cross sections in a pseudo-Nambu-Goldstone dark matter model}}, \href{https://doi.org/10.1103/PhysRevD.111.055025}{\emph{Phys. Rev. D} {\bfseries 111} (2025) 055025} [\href{https://arxiv.org/abs/2411.15755}{{\ttfamily 2411.15755}}].

\bibitem{Abe:2021nih}
S.~Abe, G.-C.~Cho and K.~Mawatari, \emph{{Probing a degenerate-scalar scenario in a pseudoscalar dark-matter model}}, \href{https://doi.org/10.1103/PhysRevD.104.035023}{\emph{Phys. Rev. D} {\bfseries 104} (2021) 035023} [\href{https://arxiv.org/abs/2101.04887}{{\ttfamily 2101.04887}}].

\bibitem{Abe:2021vat}
Y.~Abe and T.~Toma, \emph{{Direct detection of pseudo-Nambu-Goldstone dark matter with light mediator}}, \href{https://doi.org/10.1016/j.physletb.2021.136639}{\emph{Phys. Lett. B} {\bfseries 822} (2021) 136639} [\href{https://arxiv.org/abs/2108.10647}{{\ttfamily 2108.10647}}].

\bibitem{Cai:2021evx}
C.~Cai, Y.-P.~Zeng and H.-H.~Zhang, \emph{{Cancellation mechanism of dark matter direct detection in Higgs-portal and vector-portal models}}, \href{https://doi.org/10.1007/JHEP01(2022)117}{\emph{JHEP} {\bfseries 01} (2022) 117} [\href{https://arxiv.org/abs/2109.11499}{{\ttfamily 2109.11499}}].

\bibitem{Cho:2023hek}
G.-C.~Cho and C.~Idegawa, \emph{{Analyzing cancellation mechanism of the dark matter-quark scattering in a complex singlet extension of the Standard Model}}, \href{https://doi.org/10.1016/j.nuclphysb.2023.116320}{\emph{Nucl. Phys. B} {\bfseries 994} (2023) 116320} [\href{https://arxiv.org/abs/2304.10096}{{\ttfamily 2304.10096}}].

\bibitem{Maji:2023fba}
R.~Maji, W.-I.~Park and Q.~Shafi, \emph{{Gravitational waves from walls bounded by strings in SO(10) model of pseudo-Goldstone dark matter}}, \href{https://doi.org/10.1016/j.physletb.2023.138127}{\emph{Phys. Lett. B} {\bfseries 845} (2023) 138127} [\href{https://arxiv.org/abs/2305.11775}{{\ttfamily 2305.11775}}].

\bibitem{Aoki:2023tlb}
M.~Aoki and T.~Toma, \emph{{Simultaneous detection of boosted dark matter and neutrinos from the semi-annihilation at DUNE}}, \href{https://doi.org/10.1088/1475-7516/2024/02/033}{\emph{JCAP} {\bfseries 02} (2024) 033} [\href{https://arxiv.org/abs/2309.00395}{{\ttfamily 2309.00395}}].

\bibitem{Toma:2021vlw}
T.~Toma, \emph{{Distinctive signals of boosted dark matter from its semiannihilation}}, \href{https://doi.org/10.1103/PhysRevD.105.043007}{\emph{Phys. Rev. D} {\bfseries 105} (2022) 043007} [\href{https://arxiv.org/abs/2109.05911}{{\ttfamily 2109.05911}}].

\bibitem{Miyagi:2022gvy}
T.~Miyagi and T.~Toma, \emph{{Semi-annihilating dark matter coupled with Majorons}}, \href{https://doi.org/10.1007/JHEP07(2022)027}{\emph{JHEP} {\bfseries 07} (2022) 027} [\href{https://arxiv.org/abs/2201.05412}{{\ttfamily 2201.05412}}].

\bibitem{BetancourtKamenetskaia:2025noa}
B.~Betancourt~Kamenetskaia, M.~Fujiwara, A.~Ibarra and T.~Toma, \emph{{Boosted dark matter from semi-annihilations in the galactic center}}, \href{https://doi.org/10.1016/j.physletb.2025.139425}{\emph{Phys. Lett. B} {\bfseries 864} (2025) 139425} [\href{https://arxiv.org/abs/2501.12117}{{\ttfamily 2501.12117}}].

\bibitem{Georgi:2000vve}
H.~Georgi, \emph{{Lie Algebras In Particle Physics : from Isospin To Unified Theories}}, Taylor {\&} Francis, Boca Raton (2000), \href{https://doi.org/10.1201/9780429499210}{10.1201/9780429499210}.

\bibitem{Lee:1977eg}
B.W.~Lee, C.~Quigg and H.B.~Thacker, \emph{{Weak Interactions at Very High-Energies: The Role of the Higgs Boson Mass}}, \href{https://doi.org/10.1103/PhysRevD.16.1519}{\emph{Phys. Rev. D} {\bfseries 16} (1977) 1519}.

\bibitem{ATLAS:2023tkt}
{\scshape ATLAS} collaboration, \emph{{Combination of searches for invisible decays of the Higgs boson using 139 fb{\ensuremath{-}}1 of proton-proton collision data at s=13 TeV collected with the ATLAS experiment}}, \href{https://doi.org/10.1016/j.physletb.2023.137963}{\emph{Phys. Lett. B} {\bfseries 842} (2023) 137963} [\href{https://arxiv.org/abs/2301.10731}{{\ttfamily 2301.10731}}].

\bibitem{CMS:2023sdw}
{\scshape CMS} collaboration, \emph{{A search for decays of the Higgs boson to invisible particles in events with a top-antitop quark pair or a vector boson in proton-proton collisions at $\sqrt{s} = 13\,\text {Te}\hspace{-.08em}\text {V} $}}, \href{https://doi.org/10.1140/epjc/s10052-023-11952-7}{\emph{Eur. Phys. J. C} {\bfseries 83} (2023) 933} [\href{https://arxiv.org/abs/2303.01214}{{\ttfamily 2303.01214}}].

\bibitem{Hambye:2008bq}
T.~Hambye, \emph{{Hidden vector dark matter}}, \href{https://doi.org/10.1088/1126-6708/2009/01/028}{\emph{JHEP} {\bfseries 01} (2009) 028} [\href{https://arxiv.org/abs/0811.0172}{{\ttfamily 0811.0172}}].

\bibitem{DEramo:2010keq}
F.~D'Eramo and J.~Thaler, \emph{{Semi-annihilation of Dark Matter}}, \href{https://doi.org/10.1007/JHEP06(2010)109}{\emph{JHEP} {\bfseries 06} (2010) 109} [\href{https://arxiv.org/abs/1003.5912}{{\ttfamily 1003.5912}}].

\bibitem{Alloul:2013bka}
A.~Alloul, N.D.~Christensen, C.~Degrande, C.~Duhr and B.~Fuks, \emph{{FeynRules 2.0 - A complete toolbox for tree-level phenomenology}}, \href{https://doi.org/10.1016/j.cpc.2014.04.012}{\emph{Comput. Phys. Commun.} {\bfseries 185} (2014) 2250} [\href{https://arxiv.org/abs/1310.1921}{{\ttfamily 1310.1921}}].

\bibitem{Alguero:2023zol}
G.~Alguero, G.~Belanger, F.~Boudjema, S.~Chakraborti, A.~Goudelis, S.~Kraml et~al., \emph{{micrOMEGAs 6.0: N-component dark matter}}, \href{https://doi.org/10.1016/j.cpc.2024.109133}{\emph{Comput. Phys. Commun.} {\bfseries 299} (2024) 109133} [\href{https://arxiv.org/abs/2312.14894}{{\ttfamily 2312.14894}}].

\bibitem{DUNE:2020ypp}
{\scshape DUNE} collaboration, \emph{{Deep Underground Neutrino Experiment (DUNE), Far Detector Technical Design Report, Volume II: DUNE Physics}},  \href{https://arxiv.org/abs/2002.03005}{{\ttfamily 2002.03005}}.

\bibitem{DUNE:2020fgq}
{\scshape DUNE} collaboration, \emph{{Prospects for beyond the Standard Model physics searches at the Deep Underground Neutrino Experiment}}, \href{https://doi.org/10.1140/epjc/s10052-021-09007-w}{\emph{Eur. Phys. J. C} {\bfseries 81} (2021) 322} [\href{https://arxiv.org/abs/2008.12769}{{\ttfamily 2008.12769}}].

\bibitem{Paschos:2001np}
E.A.~Paschos and J.Y.~Yu, \emph{{Neutrino interactions in oscillation experiments}}, \href{https://doi.org/10.1103/PhysRevD.65.033002}{\emph{Phys. Rev. D} {\bfseries 65} (2002) 033002} [\href{https://arxiv.org/abs/hep-ph/0107261}{{\ttfamily hep-ph/0107261}}].

\bibitem{Berger:2019ttc}
J.~Berger, Y.~Cui, M.~Graham, L.~Necib, G.~Petrillo, D.~Stocks et~al., \emph{{Prospects for detecting boosted dark matter in DUNE through hadronic interactions}}, \href{https://doi.org/10.1103/PhysRevD.103.095012}{\emph{Phys. Rev. D} {\bfseries 103} (2021) 095012} [\href{https://arxiv.org/abs/1912.05558}{{\ttfamily 1912.05558}}].

\bibitem{HoefkenZink:2024hor}
J.~Hoefken~Zink, S.~Hor and M.E.~Ramirez-Quezada, \emph{{Dark matter interactions in white dwarfs: A multi-energy approach to capture mechanisms}}, \href{https://doi.org/10.1007/JHEP05(2025)160}{\emph{JHEP} {\bfseries 05} (2025) 160} [\href{https://arxiv.org/abs/2410.13908}{{\ttfamily 2410.13908}}].

\bibitem{Fermi-LAT:2015att}
{\scshape Fermi-LAT} collaboration, \emph{{Searching for Dark Matter Annihilation from Milky Way Dwarf Spheroidal Galaxies with Six Years of Fermi Large Area Telescope Data}}, \href{https://doi.org/10.1103/PhysRevLett.115.231301}{\emph{Phys. Rev. Lett.} {\bfseries 115} (2015) 231301} [\href{https://arxiv.org/abs/1503.02641}{{\ttfamily 1503.02641}}].

\bibitem{Chu:2018nki}
X.~Chu and C.~Garcia-Cely, \emph{{Core formation from self-heating dark matter}}, \href{https://doi.org/10.1088/1475-7516/2018/07/013}{\emph{JCAP} {\bfseries 07} (2018) 013} [\href{https://arxiv.org/abs/1803.09762}{{\ttfamily 1803.09762}}].

\bibitem{Kamada:2017gfc}
A.~Kamada, H.J.~Kim, H.~Kim and T.~Sekiguchi, \emph{{Self-Heating Dark Matter via Semiannihilation}}, \href{https://doi.org/10.1103/PhysRevLett.120.131802}{\emph{Phys. Rev. Lett.} {\bfseries 120} (2018) 131802} [\href{https://arxiv.org/abs/1707.09238}{{\ttfamily 1707.09238}}].

\end{thebibliography}\endgroup

\end{document}